\documentclass[journal=jcisd8,manuscript=article]{achemso}
\usepackage{titlesec}
\titleformat{\section}{\normalfont\Large\bfseries}{\thesection.}{1em}{}
\titleformat{\subsection}{\normalfont\large\bfseries}{\thesubsection.}{1em}{}

\usepackage{CJK}
\usepackage{array}
\newcolumntype{C}{>{\centering\arraybackslash}c} % 定义 C 为居中对齐
\usepackage{tabularx}
\usepackage{longtable,multirow}
\usepackage{color}
\usepackage{colortbl}
\usepackage{graphicx}
\usepackage{amsmath,amssymb,amsfonts,amsthm}
\usepackage{bm}
\usepackage{algorithmicx}
\usepackage[ruled]{algorithm}
\usepackage{algpseudocode}
\usepackage{enumerate}
\usepackage{subfigure}
\usepackage{hyperref}
\usepackage[utf8]{inputenc} % allow utf-8 input
\usepackage[T1]{fontenc}    % use 8-bit T1 fonts
\usepackage{hyperref}       % hyperlinks
\usepackage{url}            % simple URL typesetting
\usepackage{booktabs}       % professional-quality tables
\usepackage{amsfonts}       % blackboard math symbols
\usepackage{nicefrac}       % compact symbols for 1/2, etc.
\usepackage{microtype}      % microtypography
\usepackage{xcolor}         % colors
\usepackage{makecell}

\usepackage{chemformula} % Formula subscripts using \ch{}
\usepackage[T1]{fontenc} % Use modern font encodings

\author{Zicheng Ma}
\affiliation{Changping Laboratory, Beijing 102200, China.}
\author{Chuanliu Fan}
\affiliation{School of Computer Science and Technology, Soochow University, Suzhou, China}
\author{Zhicong Wang}
\affiliation{School of Computer Science and Technology, Soochow University, Suzhou, China}
\author{Zhenyu Chen}
\affiliation{Beijing National Laboratory for Molecular Sciences, College of Chemistry and Molecular Engineering, 
Peking University, Beijing 100871, China.}
\author{Xiaohan Lin}
\affiliation{Beijing National Laboratory for Molecular Sciences, College of Chemistry and Molecular Engineering, 
Peking University, Beijing 100871, China.}
\author{Yanheng Li}
\affiliation{Beijing National Laboratory for Molecular Sciences, College of Chemistry and Molecular Engineering, 
Peking University, Beijing 100871, China.}
\author{Shihao Feng}
\affiliation{Changping Laboratory, Beijing 102200, China.}
\author{Jun Zhang}
\affiliation{Changping Laboratory, Beijing 102200, China.}
\email{jzhang@cpl.ac.cn}
\author{Ziqiang Cao}
\affiliation{School of Computer Science and Technology, Soochow University, Suzhou, China}
\email{zqcao@suda.edu.cn}
\author{Yi Qin Gao}
\affiliation{Beijing National Laboratory for Molecular Sciences, College of Chemistry and Molecular Engineering, 
Peking University, Beijing 100871, China.}
\email{gaoyq@pku.edu.cn}
\title[An \textsf{achemso} demo]
  {ProtTeX: Structure-In-Context Reasoning and Editing of Proteins with Large Language Models}

\begin{document}


\begin{abstract}
Large language models have made remarkable progress in the field of molecular science, particularly in understanding and generating functional small molecules. This success is largely attributed to the effectiveness of molecular tokenization strategies. In protein science, the amino acid sequence serves as the sole tokenizer for LLMs. However, many fundamental challenges in protein science are inherently structure-dependent. The absence of structure-aware tokens significantly limits the capabilities of LLMs for comprehensive biomolecular comprehension and multimodal generation. To address these challenges, we introduce a novel framework, ProtTeX, which tokenizes the protein sequences, structures, and textual information into a unified discrete space. This innovative approach enables joint training of the LLM exclusively through the Next-Token Prediction paradigm, facilitating multimodal protein reasoning and generation. ProtTeX enables general LLMs to perceive and process protein structures through sequential text input, leverage structural information as intermediate reasoning components, and generate or manipulate structures via sequential text output. Experiments demonstrate that our model achieves significant improvements in protein function prediction, outperforming the state-of-the-art domain expert model with a twofold increase in accuracy. Our framework enables high-quality conformational generation and customizable protein design. For the first time, we demonstrate that by adopting the standard training and inference pipelines from the LLM domain, ProtTeX empowers decoder-only LLMs to effectively address diverse spectrum of protein-related tasks.
\end{abstract}

\section{Introduction}
Proteins are fundamental to a wide range of biological processes and play a critical role in cellular function and regulation. In recent years, the integration of physical modeling with advanced deep learning methodologies has revolutionized our ability to investigate the physicochemical properties and functional dynamics of proteins. This synergy has led to groundbreaking achievements, including the precise characterization of protein sequences \cite{linEvolutionaryscalePredictionAtomiclevel,hayesSimulating500Million2025}, the highly accurate prediction of protein structures \cite{jumperHighlyAccurateProtein2021b,abramsonAccurateStructurePrediction2024,baekAccuratePredictionProtein}, and the innovative design of protein sequences guided by various conditional constraints \cite{renAccurateRobustProtein2024,dauparasRobustDeepLearning,ingrahamIlluminatingProteinSpace2023,madaniLargeLanguageModels2023}. The transformative impact of artificial intelligence on protein science was further underscored by the 2024 Nobel Prize in Chemistry, which recognized pioneering advancements in protein engineering. Specifically, AI-driven tools such as AlphaFold2 \cite{jumperHighlyAccurateProtein2021b} and RFdiffusion \cite{watsonNovoDesignProtein2023b} have redefined protein structure prediction with unprecedented precision and facilitated in silico protein design. Despite these advancements, the diverse and multifaceted nature of protein-related challenges necessitates task-specific models tailored to distinct biological questions. The emergence of numerous specialized tools highlights the inherent multitask complexity of protein science, where solutions often require problem-specific architectures rather than a single unified framework. \par

Large language models (LLMs) exhibit scalability, emergent abilities, and generality, enabling them to transcend the limitations of single-task models and facilitate cross-domain knowledge transfer\cite{zhao2024surveylargelanguagemodels,wei2022emergentabilitieslargelanguage}. They are fundamentally reshaping the paradigm of scientific research and serve as a highly suitable unified framework for multitask learning. In the realm of small molecules, advanced LLMs have demonstrated remarkable capabilities in predicting molecular properties, understanding functional characteristics, and designing novel molecules. For instance, Chen et al.\cite{chen2024mattergptgenerativetransformermultiproperty} developed MatterGPT, a generative transformer model that employs the SLICES (Simplified Line-Input Crystal-Encoding System)~\cite{xiaoInvertibleInvariantCrystal2023} representation to achieve on-demand inverse design of solid-state materials with single and multiple targeted properties. The success of LLMs in this domain can be attributed to the use of domain-specific tokenizers, such as SMILES (Simplified Molecular Input Line Entry System)~\cite{smileweininger}, SELFIES (SELF-referencIng Embedded Strings)~\cite{Krenn_2020}, and SLICES, which effectively encode molecular representations for learning and inference. In the field of proteins, tokenization has been largely limited to the use of one-letter amino acid representations~\cite{wangInstructProteinAligningHuman2023a,shenTourSynbioMultiModalLarge2024a,fangMolInstructionsLargeScaleBiomolecular2024a}. For example, Llama2-molinst-protein-7B~\cite{fangMolInstructionsLargeScaleBiomolecular2024a} is fine-tuned from the Llama2-7B model using the protein-oriented dataset from Mol-instructions~\cite{fangMolInstructionsLargeScaleBiomolecular2024a}, enabling diverse protein function understanding. However, tokenizing protein sequences using one-letter abbreviations leads to ambiguity with textual characters and often results in mismatches between amino acid length and tokenized sequence length, which can obscure the semantic representation of sequence elements. Furthermore, sequence-only representations are insufficient for fully understanding proteins, as functional inference often requires structural information. To tackle this challenge, many approaches employ compositional modality-specific protein encoders~\cite{lvProLLaMAProteinLanguage2024,wangProtChatGPTUnderstandingProteins2025,luoBioMedGPTOpenMultimodal2023,xiang1FAPMFunctional,liu-etal-2024-prott3}. For example, BioMedGPT-LM-10B~\cite{luoBioMedGPTOpenMultimodal2023} employs ESM2-3B~\cite{schaferEvolutionarySelectionProteins2023} as its encoder and is fine-tuned from the Llama2-7B~\cite{touvron2023llama2openfoundation} model using millions of protein-text question-answering pairs, enabling the generation of natural language descriptions for proteins. However, these models face optimization challenges and are limited to text generation, struggling with multimodal generation and reasoning. Moreover, Several purported LLM-based models \cite{xiang1FAPMFunctional,wangInstructProteinAligningHuman2023a} still rely on classification-based protein function prediction, restricting prompt flexibility and model adaptability. In biological research, function classification is inherently open-ended due to the continuous discovery of novel proteins and the evolving nature of taxonomies. For instance, in the Gene Ontology database~\cite{10.1093/nar/gkh036}, the number of molecular function terms for humans increased by 26\% in 2018 compared to 2016\cite{10.1093/nar/gky1055}. Given LLM architectures' strengths in generative modeling, we emphasize generative capabilities over rigid classification paradigms. Overly constrained prompts and classification heads would limit the model’s ability to handle novel or ambiguous functions, undermining the goal of developing a robust and adaptive protein chat system. Current models also exhibit a significant limitation in their inability to perform reasoning, e.g. the Chain-of-Thought (CoT) approach~\cite{wei2022chain}, which introduces a sequence of intermediate reasoning steps within in-context examples, facilitating complex reasoning processes in LLMs. In protein science, understanding or annotating proteins often necessitates the derivation and inference of information from sequences, structures, and other relevant data. To enable LLMs to perform reasoning and deduction on protein structures, the introduction of an effective structure-based tokenizer is essential. However, the integration of such a tokenizer into LLMs and the exploration of reasoning abilities remain unexplored.

Inspired by the multi-modal LLMs Emu3~\cite{wangEmu3NextTokenPrediction2024} and Chamelon~\cite{teamChameleonMixedModalEarlyFusion2024a}, we present ProtTeX, a dedicated framework for tokenizing both the 1D sequence and 3D structure of proteins for LLMs, akin to SMILES. Just as TeX provides precise control over document layout and formatting, ProtTeX empowers researchers for protein data formatting and editing, which enables LLMs to reason and generate sequences that arbitrarily interleave textual and protein modality. ProtTeX can employ the advanced Reasoning paradigm, CoT reasoning, to enhance LLMs' capability in protein-related tasks. This approach allows structural or textual information to serve as a key logical component in the reasoning process, making protein problem solving more transparent, logical, interpretable, and controllable. We have constructed a unified foundational framework that leverages a single model and loss function based solely on the next-token prediction (NTP) strategy, enabling seamless adaptation to diverse protein-related downstream tasks. In comparison to the state-of-the-art domain expert LLMs BiomedGPT~\cite{luoBioMedGPTOpenMultimodal2023} and Llama2-molinst-protein-7B~\cite{fangMolInstructionsLargeScaleBiomolecular2024a}, our approach demonstrates a twofold enhancement in accuracy for protein function prediction. Our findings reveal that the CoT approach and sampling strategies from the field of LLM can be effortlessly adapted to the protein domain, enabling high-accuracy protein structure prediction and conformational sampling. Additionally, the integration of arbitrary textual inputs allows for controllable protein generation based on human-defined prompts. For the first time, we demonstrate that decoder-only LLMs are capable of understanding, predicting, and designing proteins.

\section{Methods}
\begin{figure*}[t]
  \centering
  \includegraphics[width=0.7\linewidth]{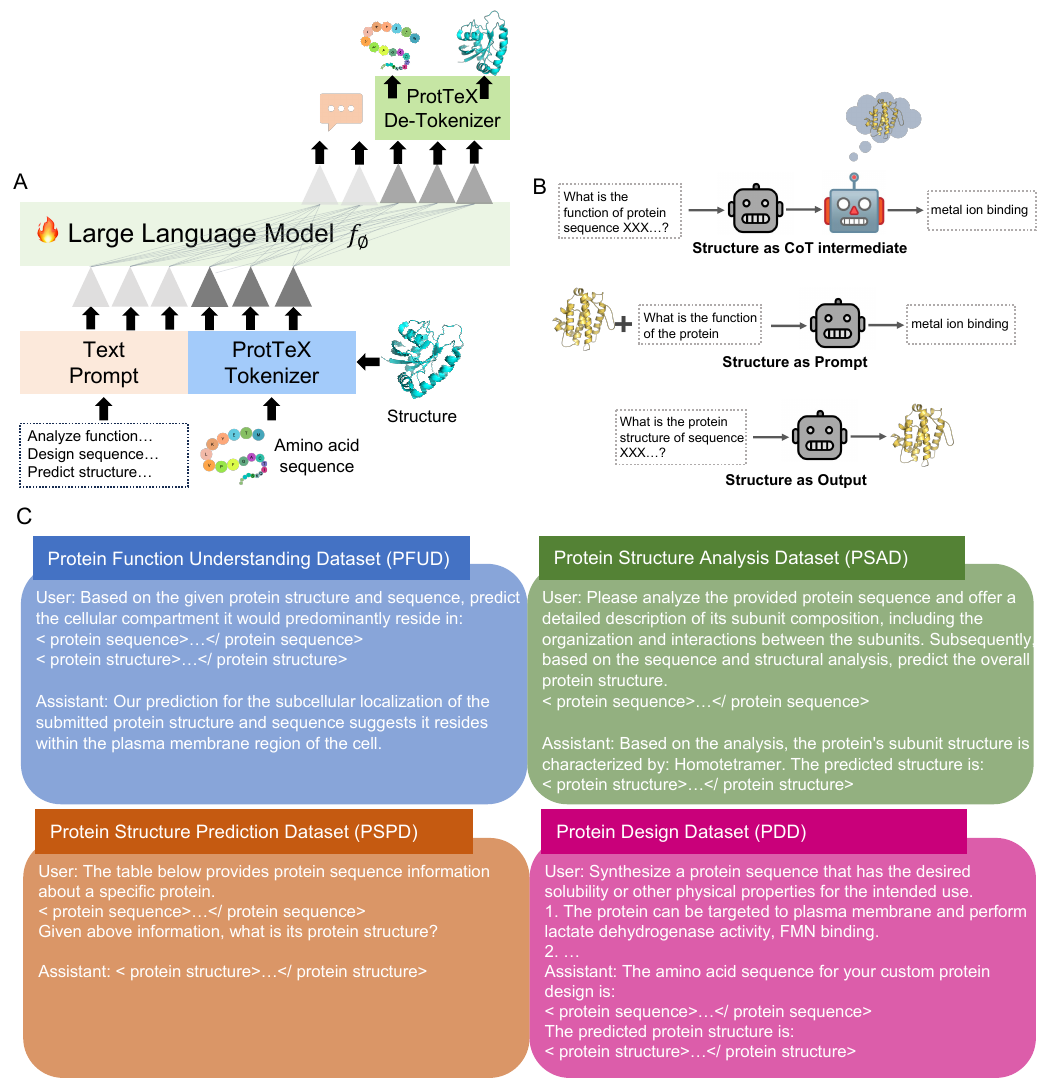} % 调整缩放比例
  \caption{(A) Overview of model architecture. (B) Structure-In-Context schematic diagram. The model enables the protein structures as input, output or CoT intermediate. (C) Prompt template of different dataset.}
  \label{fig:model}
\end{figure*}

ProtTeX represents protein sequences, structures, and natural language text as a series of reversible discrete tokens, leveraging the unified training paradigm of auto-regressive transformers\cite{kaplan2020scalinglawsneurallanguage,10.5555/3295222.3295349}. During training, we construct various prompts with arbitrary orders of protein and text, enabling the model to complete different downstream tasks, ranging from unimodal inference to multimodal CoT generation. The main architechture of the model is shown in Figure ~\ref{fig:model}A.  

\subsection{Tokenizing All-Atom Protein Structures} \label{tokenizer}
ProtTeX tokenizes the protein backbone structure following the work of Lin et al.~\cite{zhangProTokensProbabilisticVocabulary}. Here, we outline the main technical strategy. The orginal work  focuses on tokenizing the metastable conformational structure of $x$ of proteins into discrete tokens through a model that includes three main components: the encoder $f_{\theta}$, the tokenizer $h_{\theta}$, and the decoder $g_{\phi}$. The original training objective is simply to reconstruct the encoded structures.

\begin{align}
    g_{\phi}(h_{\theta}(f_{\theta}(x)))\approx x \label{protoken_1}
\end{align}

The encoder $f_{\theta}$ is a parameterized SE(3)-invariant module that transforms a protein structure $x$ with $N_{res}$ residues into a $d_{s}$-dimensional single representation $s \in  \mathbb{R}^{N_{res} \times d_s}$ and a  $d_{p}$-dimensional pair representation $p \in  \mathbb{R}^{N_{res}^2 \times d_p}$. We modified the EvoFormer and structure module framework in AlphaFold2~\cite{jumperHighlyAccurateProtein2021b} to develop a "sandwich-like" transformer module that updates both single and pairwise representations and finally outputs a $d$-dimensional representation $f_{\theta,r}(x) \in \mathbb{R}^d$ for each residue $r(1 \leq r \leq N_{res})$. 

The tokenizer $h_{\theta}$ utilizes vector quantization~\cite{oordNeuralDiscreteRepresentation2018,zhengOnlineClusteredCodebook2023} techniques commonly used in image tokenization. Specifically, we initialize a codebook with 512 codes, $\{c_i\},c_i \in \mathbb{R_d}$, each input vector $f_{\theta,r}(x)$ is assigned to the nearest code $c_i$ in the code book via a nearest neighbor search:
\begin{align}
    z_{x,r} = \operatorname{argmin}_{c_i \in \{c_i\}} \|f_{\theta,r}(x) - c_i  \|
\end{align}
The tokenized word for residue $r$ is defined as the code $z_{x,r} = c_i$ and the corresponding "token index" $i$, which will be used in LLMs. 
The decoder is similarly an SE(3)-equivariant "sandwich-like" transformer that samples protein structures from a metastable ensemble corresponding to a given tokenized  string. We apply an alignment loss and a uniformity loss to optimize the embedding space~\cite{10.5555/3524938.3525859}. For further details, refer to the original paper~\cite{zhangProTokensProbabilisticVocabulary}. 

Then, we use the original amino acid sequence as our protein side-chain tokenizer. Instead of directly using abbreviation letters or other compositional encoders, we simply add 20 new special tokens to LLMs to represent the protein sequence, similar to Emu3\cite{wangEmu3NextTokenPrediction2024}. We reinitialize 512+20=532 new tokens using the same methodology as that applied to the original textual tokens.

\subsection{Task-Unifying Prompts for Structure-In-Context Learning}

Inspired by Chameleon~\cite{team2024chameleon}, which constructed any ordering of images and text during training from text-only, to single text/image pairs to full interleaved text-image documents, we also construct interleaved protein-text QA questions. Specifically, we incorporate special tokens to merge protein sequences, protein structures, and natural language prompts, creating document-like inputs for the training process. The resulting training data are structured as follows.

\begin{center}
[BOS]\{question or description text\}[EOS]\\
< protein sequence>\{sequence tokens\}</ protein sequence> \\
< protein structure>\{structure tokens\}</ protein structure>  \\
\end{center}

Here, BOS and EOS are the original special tokens in the LLM tokenizer. The order of the three modalities above can be arbitrarily changed depending on different downstream tasks. 

Our token-based framework enables us to construct the CoT-like prompt template. As shown in Figure~\ref{fig:model}C, four prompt templates are designed to support the multimodal CoT reasoning process of the model. This framework facilitates the model  in using protein structures as input, output, or CoT intermediates, as shown in Figure~\ref{fig:model}B and Figure~\ref{fig:model}C. Specifically, the model can first generate a descriptive explanation of structure based on sequence, then produce the corresponding structure. Alternatively, it can generate a structure from a sequence and subsequently derive a description of the protein using both sequence and structure. The corresponding datasets are introduced in Section~\ref{Dataset}. 

In this paper, we focus primarily on the tasks of protein function prediction, protein structure generation, and controllable protein design. This prompt construction framework also enables researchers to explore other interesting tasks, such as inverse folding or structure design, requiring only fine-tuning tailored to specific objectives. We support any-to-any modality transformation, serving as a general and reliable foundational framework.

% \begin{figure*}[t]
%   \centering
%   \includegraphics[width=0.7\linewidth]{Figure/Figure_2.pdf} % 调整缩放比例
%   \caption{Prompt Template}
%   \label{fig:prompt}
% \end{figure*}

\subsection{Data and Models}
%蒸馏数据集来源,psp: https://arxiv.org/pdf/2206.12240

\subsubsection{Dataset}\label{Dataset}
We first curated a database of 3.36 million proteins, including their sequences and structures, from the clustered AlphaFold Protein Structure Database (AFDB) v4 dataset~\cite{hernandezClusteringPredictedStructures}, the Swiss-Prot database released in May 2022~\cite{10.1093/nar/28.1.45}, and RCSB PDB~\cite{10.1093/nar/28.1.235}. All of the proteins were released before July 25, 2022. We then processed this database using ProtTeX introduced in Section~\ref{tokenizer}, for structural reconstruction and filtering to obtain the sequence and structural tokens for every protein. The dataset is subsequently split into training (90\%), validation (5\%), and test (5\%) sets.

The protein-related QA pairs were curated from Mol-Instruction\cite{fangMolInstructionsLargeScaleBiomolecular2024a} and ProteinLMBench\cite{shenFinetuningDatasetBenchmark2024}, with all protein accessions sourced from UniProt~\cite{10.1093/nar/gkae1010}. By mapping the corresponding accessions of specific QA pairs from the training, validation, and test sets, we constructed three distinct datasets: Protein Function Understanding Dataset (PFUD), Protein Structure Analysis Dataset (PSAD), and Protein Design Dataset (PDD). Among these, PFUD and PDD were derived and modified from the Mol-Instruction dataset, while PSAD was derived from a subset of ProteinLMBench. If an accession does not exist in our database, the corresponding questions were dropped. The remaining proteins without corresponding QA pairs formed the Protein Structure Prediction Dataset (PSPD). The total dataset for training our main model is composed of the aforementioned four components—PFUD, PSAD, PDD, and PSPD—and is randomly shuffled at the beginning of each epoch to ensure robust training and prevent any potential bias introduced by the order of the data. The token counts for each subset are presented in Table~\ref{tab:dataset}. For more details on our dataset, please see the supplementary section~\ref{supp_dataset}.

\begin{table}[ht]
    \centering
    \renewcommand{\arraystretch}{1.25}
    \caption{Fine-Tuning Dataset Statistics}
    \setlength{\belowcaptionskip}{12pt}
    \begin{tabular}{l|rr }
        \hline
        Dataset & \# of Samples & \# of Tokens \\
        \hline
        PFUD & 429,201 & 320.4M  \\
        PDD & 192,617 & 146.8M \\
        PSAD & 264,370 & 205.0M \\
        PSPD & 2,821,238 & 1787.8M \\
        \hline
    \end{tabular}
    \label{tab:dataset} 
\end{table}

\subsubsection{Training}
Since protein sequences and structures are fully converted into discrete tokens, we only need to train using the next-token prediction task with the standard cross-entropy loss. Given a sequence of tokens $\mathbf{x} = (x_1, x_2, \dots, x_T)$, the auto-regressive model predicts the probability of each token $x_t$ conditioned on the previous tokens $\mathbf{x}_{<t} = (x_1, x_2, \dots, x_{t-1})$. The loss function $\mathcal{L}$ is defined as the negative log-likelihood of the sequence:
\begin{align}
\mathcal{L}(\mathbf{x}) = -\sum_{t=1}^{T} \log P(x_t \mid \mathbf{x}_{<t}; \theta)
\end{align}
Instead of training a completely new model from scratch, we opt for continuous pre-training~\cite{ke2023continualpretraininglanguagemodels} and supervised fine-tuning~\cite{jiang2024supervisedfinetuningturnimproves} of existing general LLMs. We assign equal weight to the tokens of the protein modality and natural language, given the critical importance of protein information. For further details on training, please refer to the supplementary section~\ref{training_details}.

% While this approach may lead to catastrophic forgetting~\cite{Kirkpatrick_2017} in the model, we have chosen to prioritize the model's specialized utility over addressing this issue for the time being. We are confident that researchers will develop excellent solutions to mitigate catastrophic forgetting during domain knowledge integration in the future.

\subsubsection{Inference and sampling}\label{sampling}
In the reasoning process of LLMs, the sampling strategy plays a pivotal role. For different downstream tasks, we adopt different sampling strategies. We employ simple greedy search for all protein understanding tasks. For protein structure analysis or prediction task, a novel sampling strategy, \textbf{Beam Search with Lowest perplexity (PPL)}, is designed to enhance the applicability of autoregressive models. Let us formally recall the perplexity metric of the output $Y$ given a specific prompt $p$:

\begin{equation} \mathcal{P}(Y|p) = \exp\left( -\frac{1}{n} \sum_{t=m+1}^{m+n} \log p(y_t| y_{<t},\theta) \right) \end{equation}

The sampling strategy formalizes the generation process as follows:

\begin{equation} \hat{y} = \underset{y \in \mathcal{B}}{\arg\min} \mathcal{P}(y|p) \end{equation}

where B denotes the beam search space defined by:

\begin{equation} \mathcal{B} = {y^{(1)},...,y^{(k)}} \sim p_{\theta}(y|p) \end{equation}

The nucleus sampling strategy~\cite{holtzman2020curiouscaseneuraltext} is used for the multi-conformation generation and protein design tasks.

\subsubsection{Baseline Setups}\label{baselines}
To systematically evaluate the performance of our model in protein understanding tasks, we introduce the following baselines.
\begin{itemize}
    \item \textbf{Llama3-Instruct} Meta-Llama-3-8B-Instruct~\cite{grattafioriLlama3Herd2024} is an advanced language model developed by Meta, featuring 8 billion parameters and optimized for instructional fine-tuning to enhance conversational performance. Despite its relatively smaller parameter size compared to other models, the Meta-Llama-3-8B-Instruct has demonstrated superior performance, even surpassing the 70-billion-parameter Llama-2 model~\cite{touvron2023llama2openfoundation} in various benchmarks.

    \item \textbf{BioMedGPT-LM-10B~\cite{luoBioMedGPTOpenMultimodal2023}} BioMedGPT is a domain-specific LLM fine-tuned on a large selection of biological scientific corpora, including protein-related questions. The model encodes the protein sequence with ESM-3B and uses BioMedGPT-LM-7B as the decoder to generate responses.

    \item \textbf{Llama2-molinst-protein-7B} This model was developed by the research group of the Mol-Instructions~\cite{fangMolInstructionsLargeScaleBiomolecular2024a} dataset. They performed full-parameter fine-tuning of Llama2-7B~\cite{touvron2023llama2openfoundation} using the protein-oriented dataset in Mol-Instructions. For our inference, identical parameters to those specified in the official scripts were employed.

    \item \textbf{Llama3-AAseq-FT} Supervised fine-tuned on the Meta-Llama-3-8B-Base model using textual protein sequence information with our PFUD dataset.

    \item \textbf{ProtT3-FT} ProtT3~\cite{liu-etal-2024-prott3} empowers an LLM to understand protein sequences by incorporating ESM-3B~\cite{schaferEvolutionarySelectionProteins2023} as its protein understanding module, enabling effective protein-to-text generation. Although it has undergone large-scale pretraining and fine-tuning on protein-text retrieval and generation, it imposes rigid constraints on its question templates, lacking support for diverse prompt variations and questions, which represents a significant limitation and undermines its claim as a genuine multimodal large model. To address this critical issue, we fine-tuned the model using our PFUD dataset and selected the best model in benchmark evaluations, which resulted in substantial performance improvements and enhanced model capabilities.

    \item \textbf{$\text{ProtTeX}_{\text{Llama3}}$} Our main proposed model, which is supervised fine-tuned on Meta-Llama-3-8B-Base using the ProtTeX tokenizer with our total dataset, which incorporates training data from different downstream tasks. The textual protein sequence information in PFUD dataset is replaced by our tokenized protein sequence and structure. We also conducted ablation experiments under different training conditions, including various training strategies such as Low-Rank Adaptation (LoRA)~\cite{hu2021loralowrankadaptationlarge}, different training sets, and model scales. For more details on the ablation experiments, please refer to Table~\ref{tab:result_ablation}, Table~\ref{tab:ablation_configuration}, and Section~\ref{supp_ablation}.
    
\end{itemize}

To ensure a fair comparison between models, we use the same question prompts for all models except BiomedGPT. Given that our training prompts are designed using the Mol-Instructions template, which exhibits considerable diversity in phrasing, we systematically transform all prompts into BiomedGPT's preferred format when conducting BiomedGPT inference. The format is \textit{What is [specific attribute] of this protein?}. This standardization is implemented to eliminate potential bias arising from prompt formulation differences while maintaining the essential query content.

\subsubsection{Metrics}\label{metirc}
Traditional multi-class classification models for protein function prediction rely on the CAFA~\cite{radivojacLargescaleEvaluationComputational2013} evaluation metrics. These metrics require the model to output scores for each classification head to compute the Fmax. However, these metrics are not applicable to large language generation models, as we do not employ a classification-based approach and thus do not output scores for individual classes. Consequently, we have developed a novel evaluation framework tailored to these models. We evaluate the model's output based on two key aspects: fluency and domain-specific accuracy. First, we employ two classical metrics, BLEU~\cite{papineniBLEUMethodAutomatic2001} and ROUGE~\cite{linROUGEPackageAutomatic}, which are widely used for evaluating machine translation quality. These metrics measure the overlap between machine-generated text and reference translations by comparing n-grams. 

Second,  we propose the Exact Match Jaccard Index (EMJI), a novel metric designed to evaluate the degree of overlap between the biological keywords present in the ground truth labels and those generated by the model. The JI Index score for a single question $s$ is defined as:

\begin{equation}
\text{JI}(s) = \frac{|K_{label} \cap K_{pred}|}{|K_{label} \cup K_{pred}|}
\end{equation}

% \begin{equation} EM(s) = \begin{cases} 1, & \text{if all the keywords in label are covered by } s\\
% 0, & \text{otherwise} \end{cases} \end{equation}

where $K_{label}$ represents the set of biological keywords extracted from the ground truth label of question $s$. $K_{pred}$ denotes the set of biological keywords extracted from the model's output of question $s$. The overall EMJI reported is then computed as the average of JS Index across all questions in the test set:

\begin{equation} \text{EMJI} = \frac{1}{N} \sum_{i=1}^{N} \text{JI}(s_i) \end{equation}

where N is the total number of questions in the test set.

Due to the diversity of labels in the original dataset and generated output, manually extracting keywords is highly challenging. To address this, we use DeepSeek-V3~\cite{deepseekai2025deepseekr1incentivizingreasoningcapability} to automatically extract keywords and perform exact matching using in-context learning, thereby computing the EMJI. This approach ensures a more efficient and consistent evaluation process while maintaining high precision in keyword matching.

\section{Results and discussion}

\subsection{ProtTeX Enables Structure-In-Context Protein Understanding}
For our comparative analysis, we curate single-turn dialogue samples from the PFUD test set, resulting in a collection of 5,836 distinct protein queries. These single-turn dialogues encompass six distinct domains of inquiry pertaining to proteins: molecular function, subcellular location, biological process, domains or motifs, overview of features, and multi-attribute. Specifically, the "overview of features" domain prompts the model to provide a concise summary of the most critical functional aspects of the protein, which may encompass elements of both molecular function and biological process. Within the "multi-attribute" domain, our prompts are designed to query combinations of two or three attributes from the preceding five domains, requiring the model to address them collectively in its response. We select three open-source models for our benchmark: BioMedGPT-LM-10B (BioMedGPT)~\cite{luoBioMedGPTOpenMultimodal2023}, a domain-specific expert model finetuned on protein-related questions. Llama2-molinst-protein-7B, a state-of-the-art protein understanding LLM fine-tuned on Mol-Instructions. Llama3-Instruct (8B)~\cite{grattafioriLlama3Herd2024}, a general-purpose instruction tuned LLM. To compare our model with other compositional models, we also fine-tuned a state-of-the-art compositional protein understanding model, ProtT3~\cite{liu-etal-2024-prott3}, as the original ProtT3 constrains the prompts and could not be adapted to other questions. For more details on the models, please refer to the baseline sections~\ref{baselines}.

\begin{figure*}[t]
  \centering
  \includegraphics[width=1.0\linewidth]{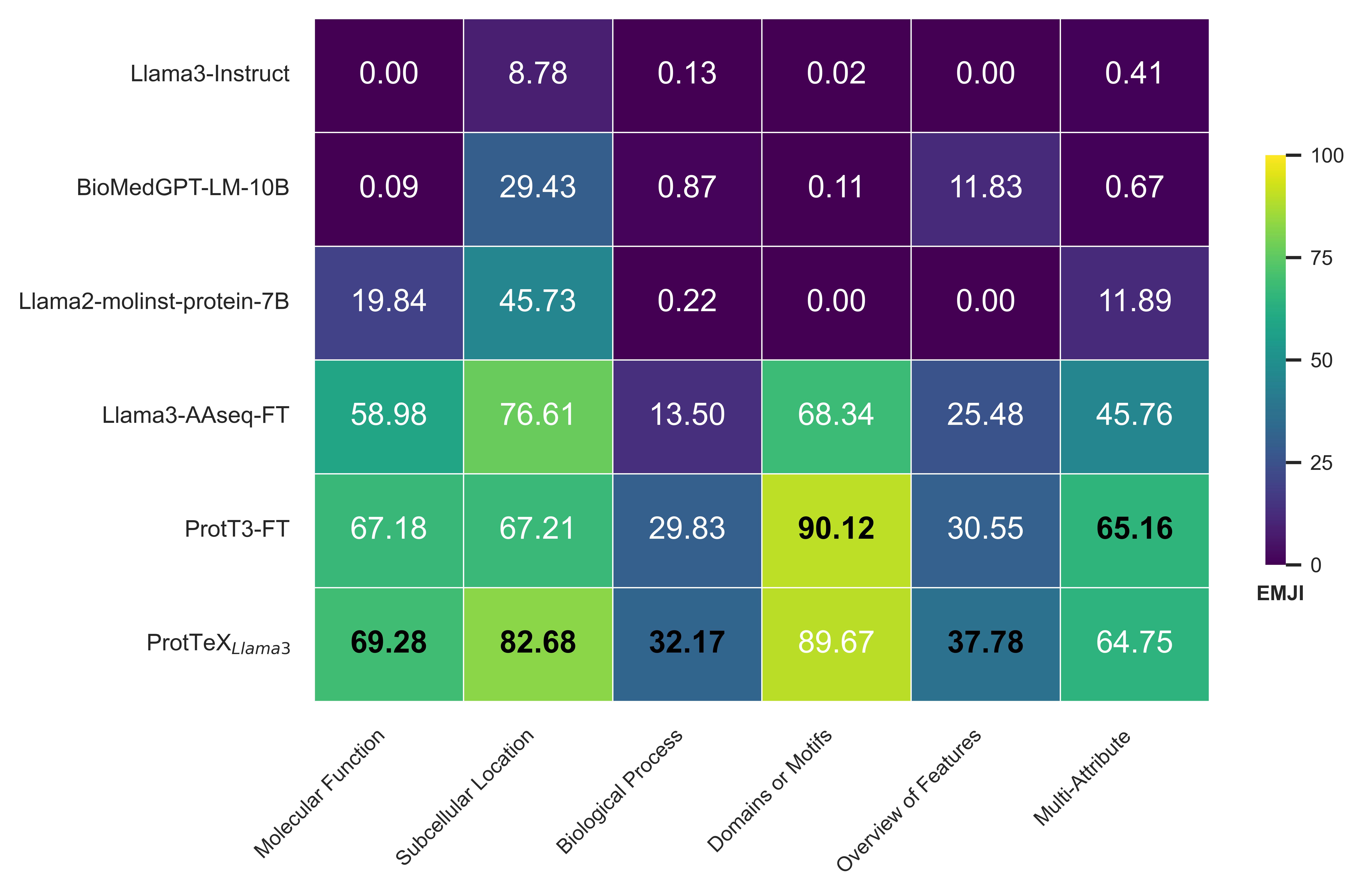} % 调整缩放比例
  \caption{Heatmap illustrates the Exact Match Jaccard Index (EMJI) of various models across different protein understanding tasks in the PFUD test set, including Molecular Function (n=1,127), Subcellular Location (n=2,071), Biological Process (n=459), Domains or Motifs (n=886), and Multi-Attribute (n=974). The best-performing metric for each task is highlighted in bold.}
  \label{fig:understanding}
\end{figure*}

As shown in Table~\ref{tab:result1}, our analysis demonstrates that the multi-task fine-tuned model $\text{ProtTeX}_{\text{Llama3}}$ achieves optimal performance, excelling in both linguistic fluency and accuracy in addressing domain-specific professional protein queries, highlighting the effectiveness of our ProtTeX framework and the existence of overlapping subspaces among different tasks which facilitate mutual enhancement during the training process. Furthermore, we conduct a comparative analysis by fine-tuning models exclusively on the PFUD dataset to evaluate their respective performance in the ablation study~\ref{supp_ablation}. The experimental results in Table~\ref{tab:result1} and Table~\ref{tab:result_ablation} demonstrate that incorporating the novel ProtTeX tokenizer significantly outperforms the the fine-tuned models that use one-letter amino acid abbreviations (Llama2-molinst-protein-7B,Llama3-AAseq-FT) or compositional protein encoders (ProtT3-FT, BiomedGPT). These results highlights the limitations of current models in functional understanding when relying solely on textual sequence alphabets or other compositional approaches, further substantiating the beneficial impact of incorporating structural tokens in enhancing protein function comprehension. On our PFUD test set, where all QA pairs are derived exclusively from Swiss-Prot proteins, the ROUGE-L score of the Llama2-molinst-protein-7B~\cite{fangMolInstructionsLargeScaleBiomolecular2024a} model closely matches its officially reported metrics. However, we find that ROUGE-L is not a robust indicator for assessing the precision of key terms, as demonstrated by the substantially lower exact match performance. Our ablation studies presented in Table~\ref{tab:result_ablation} demonstrate that pre-training without explicit functional information significantly enhances functional comprehension. These results collectively suggest a symbiotic relationship between multimodal understanding and generation. Specifically, enhancing comprehension data improves generative task performance, and conversely, expanding generative training data strengthens interpretative capabilities. This phenomenon aligns with recent observations in computer vision research~\cite{wu2025liquidlanguagemodelsscalable}. Additionally, we evaluated domain-specific response performance, as illustrated in Figure~\ref{fig:understanding}. Although the fine-tuned ProtT3 model achieves comparable performance with our proposed model in motif recognition and multi-attribute tasks, ProtTeX consistently outperforms competing approaches across various domains. Notably, the BiomedGPT model exhibits suboptimal performance throughout our experiments. Despite extensive efforts to optimize the prompt formulation by adopting BiomedGPT’s preferred structure, the model demonstrates substantial limitations in addressing a wide range of protein-related queries. This suggests that BiomedGPT's training may be insufficient for comprehensive protein-related tasks. Moreover, approximately 2\% of BiomedGPT's responses yielded \textit{unknown} answers, indicating that the external encoder integration strategy may cause substantial representational shifts in response to sequence similarity variations, potentially increasing perplexity and reducing overall performance. These observations underscore the effectiveness of our early fusion training strategy~\cite{sun2023emu,sun2024generative,yu2023scaling}, which integrates interleaved textual and protein modalities into a unified representation, enabling natural and intrinsic connections between different modalities.

\begin{table*}[!ht]
    \centering
    \renewcommand{\arraystretch}{1.25}
    \begin{tabular}{l|c|c|c|c|C}
    \toprule
        \textbf{Model} &\textbf{EMJI} & \textbf{Bleu-2} & \textbf{Rouge-1} & \textbf{Rouge-2} & \textbf{Rouge-L}    \\ 
        \midrule
        Llama3-Instruct & 3.20 & 2.08 & 15.91 & 2.67 & 5.81   \\
        BioMedGPT-LM-10B & 11.31 & 2.41 & 18.91 & 2.99 & 14.89  \\
        Llama2-molinst-protein-7B & 22.06 & 26.25  & 45.24 & 23.47 & 38.15  \\
        Llama3-AAseq-FT & 59.04 & 37.10 & 59.65 & 37.72 & 52.50  \\
        ProtT3-FT & 65.40 & 40.79 & 61.97 & 42.53 & 56.98\\
        \midrule
        $\text{ProtTeX}_{\text{Llama3}}$  & \textbf{71.73}  & \textbf{41.54} & \textbf{63.46} & \textbf{43.17} & \textbf{57.89} \\
        \bottomrule
    \end{tabular}
    \caption{Result in PFUD test set, the best performances are marked in bold.}
    \label{tab:result1}
\end{table*}

% \begin{table*}[ht]
%     \centering
%     \scriptsize
%     \renewcommand{\arraystretch}{1.25}
%     \begin{tabular}{l|cc|ccc|ccc}
%         \toprule
%         % \textbf{Model Name} & \textbf{Scale} & \textbf{Tokenizer} & \textbf{PT} & \textbf{Dataset} & \textbf{Method} & \textbf{FT} & \textbf{Dataset} & \textbf{Method} \\
%         \textbf{Model Name} & \textbf{Tokenizer} & \textbf{Scale}  & \textbf{PT} & \textbf{Dataset} & \textbf{Method} & \textbf{FT} & \textbf{Dataset} & \textbf{Method} \\
%         \midrule
%         Llama3-AAseq-FT& AAseq & 8B  & - & - & - & \checkmark & PFUD & full \\
%         Llama3-ProtTeX-FT 1B & ProtTeX & 1B & - & - & - & \checkmark & PFUD & full \\
%         % Llama3-ProtTeX-FT & 8B & ProtTeX & - & - & - & \checkmark & PFUD & full \\
%         Llama3-ProtTeX-FT w/ PT & ProtTeX & 8B  & \checkmark & PSPD & full & \checkmark & PFUD & full \\
%         Llama3-ProtTeX-FT lora & ProtTeX & 8B  & - & - & - & \checkmark & PFUD & lora \\
%         Llama3-ProtTeX (Proposed) & ProtTeX & 8B  & - & - & - & \checkmark & PFUD & full \\
%         \midrule
%         Llama3-ProtTeX (+ tasks) & ProtTeX & 8B  & - & - & - & \checkmark & \makecell{PFUD,PSPD, \\ PDD, PSAD} & full \\
%         \bottomrule
%     \end{tabular}
%     \caption{Ablation configuration.}
%     \label{tab:ablation_configuration}
% \end{table*}

\begin{table*}[ht]
    \centering
    % \scriptsize
    \small
    \renewcommand{\arraystretch}{1.25}
    \begin{tabular}{l|c|c|c|c|c}
        \toprule
        \textbf{Model Name} & \textbf{Tokenizer} & \textbf{Scale}  & \textbf{PT} & \textbf{FT} & \textbf{Method} \\
        \midrule
        Llama3-AAseq-FT & AAseq & 8B  & - &  PFUD & Full  \\
        $\text{ProtTeX}_{\text{Llama3}}$ (w/o Multi-Task) 1B & ProtTeX & 1B & -  &  PFUD & Full \\
        $\text{ProtTeX}_{\text{Llama3}}$ (w/o Multi-Task) lora & ProtTeX & 8B  & -  & PFUD & Lora \\
        $\text{ProtTeX}_{\text{Llama3}}$ (w/o Multi-Task)  & ProtTeX & 8B  & - &  PFUD &Full \\
        $\text{ProtTeX}_{\text{Llama3}}$ (w/o Multi-Task) w/ PT & ProtTeX & 8B  &PSPD &  PFUD  & Full \\
        \midrule
        $\text{ProtTeX}_{\text{Llama3}}$ (Proposed) & ProtTeX & 8B  & -  & \makecell{PFUD,PSPD, \\ PDD, PSAD}  & Full \\
        \bottomrule
    \end{tabular}
    \caption{Ablation configuration.}
    \label{tab:ablation_configuration}
\end{table*}

\begin{table*}[!ht]
    \centering
    \small
    \renewcommand{\arraystretch}{1.25}
    \begin{tabular}{l|c|c|c|c|C}
    \toprule
        \textbf{Model} &\textbf{EMJI}& \textbf{Bleu-2} & \textbf{Rouge-1} & \textbf{Rouge-2} & \textbf{Rouge-L}    \\ 
        \midrule
        Llama3-AAseq-FT & 59.04 & 37.64 & 60.44 & 37.91 & 52.79  \\
        $\text{ProtTeX}_{\text{Llama3}}$ (w/o Multi-Task) 1B    & 64.97 & 39.40 & 61.25 & 40.56 & 55.09  \\
        $\text{ProtTeX}_{\text{Llama3}}$ (w/o Multi-Task) lora & 62.13 & 38.80 & 60.38 & 39.60 & 53.72  \\
        $\text{ProtTeX}_{\text{Llama3}}$ (w/o Multi-Task)  & 66.12 & 40.01 & 62.87 & 41.71 & 56.15 \\
        $\text{ProtTeX}_{\text{Llama3}}$ (w/o Multi-Task) w/ PT  & 70.57  & 40.39 & \textbf{63.77} & 42.69 & 57.17 \\
        \midrule
        $\text{ProtTeX}_{\text{Llama3}}$ (Proposed) & \textbf{71.73}  & \textbf{41.54} & 63.46 & \textbf{43.17} & \textbf{57.89}\\
    \bottomrule
    \end{tabular}
    \caption{Ablation study in PFUD dataset, the best performances are marked in bold.}
    \label{tab:result_ablation}
\end{table*}

\subsection{ProtTeX Enables Structure-Involved Reasoning for Proteins}

LLMs have demonstrated remarkable reasoning capabilities, particularly through Chain-of-Thought (CoT) reasoning, which involves decomposing complex problems into sequential and logical steps. This method enables models to generate coherent and contextually relevant responses. CoT reasoning has been successfully implemented in several prominent LLMs, including OpenAI-o1~\cite{openai2024openaio1card} and DeepSeek-R1~\cite{deepseekai2025deepseekr1incentivizingreasoningcapability}. In multimodal settings, Chain-of-Thought reasoning allows models to effectively integrate and analyze diverse data types—such as text, images, and audio—by establishing meaningful connections among them. Although recent studies~\cite{zhang2023multimodal,guo2025generateimagescotlets} have begun exploring the integration of CoT reasoning into autoregressive image generation, the potential applications of such reasoning techniques in the biological sciences remain largely unexplored.
 
\begin{figure*}[t]
  \centering
  \includegraphics[width=1.0\linewidth]{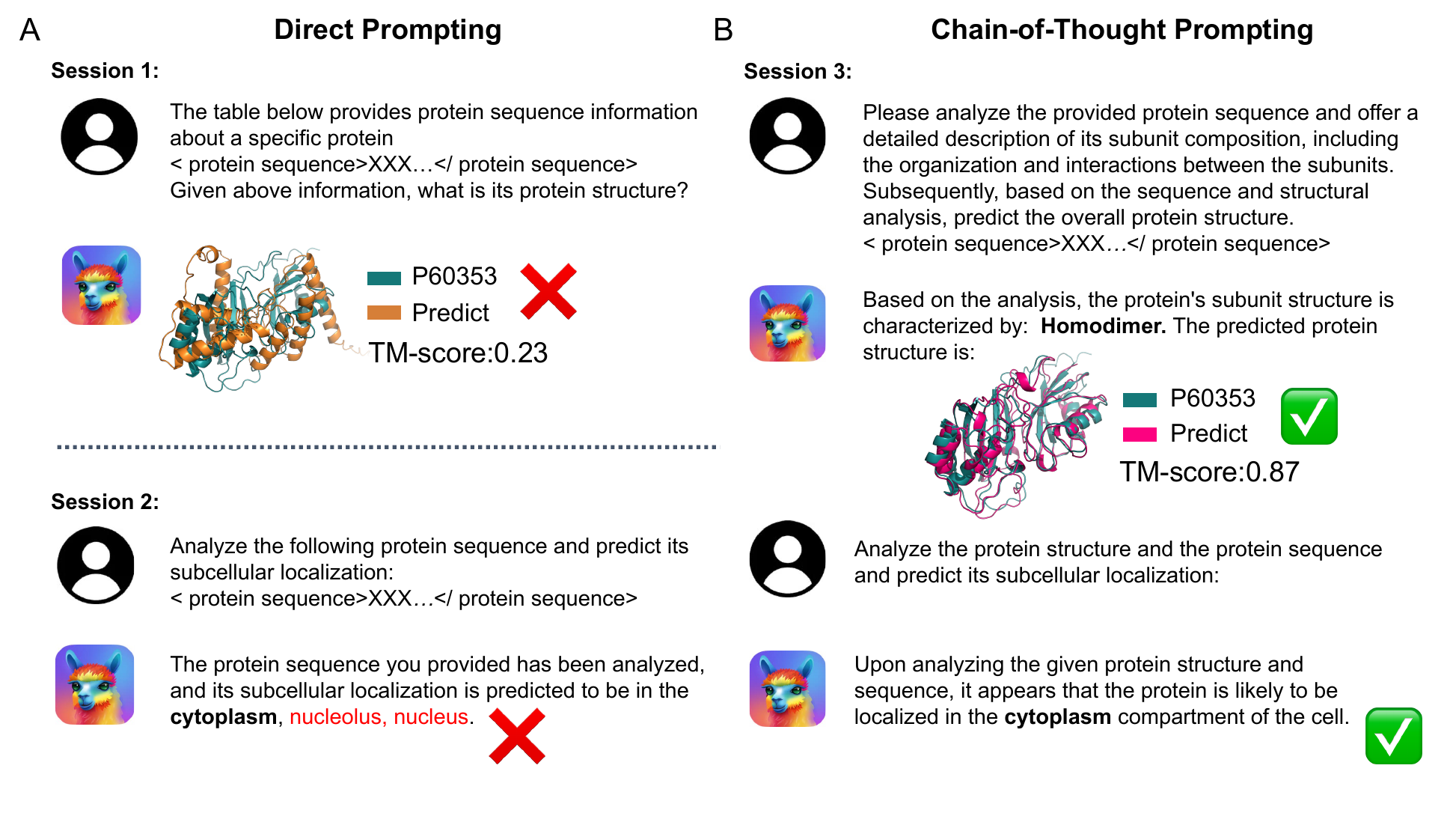} 
  \caption {Multimodal chain-of-thought with multi-round chat. (A) Direct Prompting, direct asking the question of protein structure or protein function. (B) Chain-of-Thought Prompting, first analyzes the sequence, then generates the structure, and subsequently infers the function step-by-step. The Llama icon is sourced from \url{https://github.com/alexrozanski/LlamaChat}.}
  \label{fig:chat}
\end{figure*}

During the training phase, we incorporate a small subset of CoT-like data, enabling the model to acquire CoT reasoning capabilities across modalities. Specifically, our framework can employ a step-by-step generative process: first, the model generates a protein structure analysis based on instructions provided in the protein sequence; next, it synthesizes a protein structure guided by the initial prompts and generated descriptions; finally, it produces corresponding functional textual descriptions conditioned on both the synthesized structures and previous descriptions. This workflow establishes a prototype for multi-round multimodal reasoning, as illustrated in Figure~\ref{fig:chat}. Additionally, the bidirectional protein-text generation mechanism significantly enhances the model's cross-modal understanding and reasoning capabilities.

\begin{figure*}[!ht]
  \centering
  \includegraphics[width=0.8\linewidth]{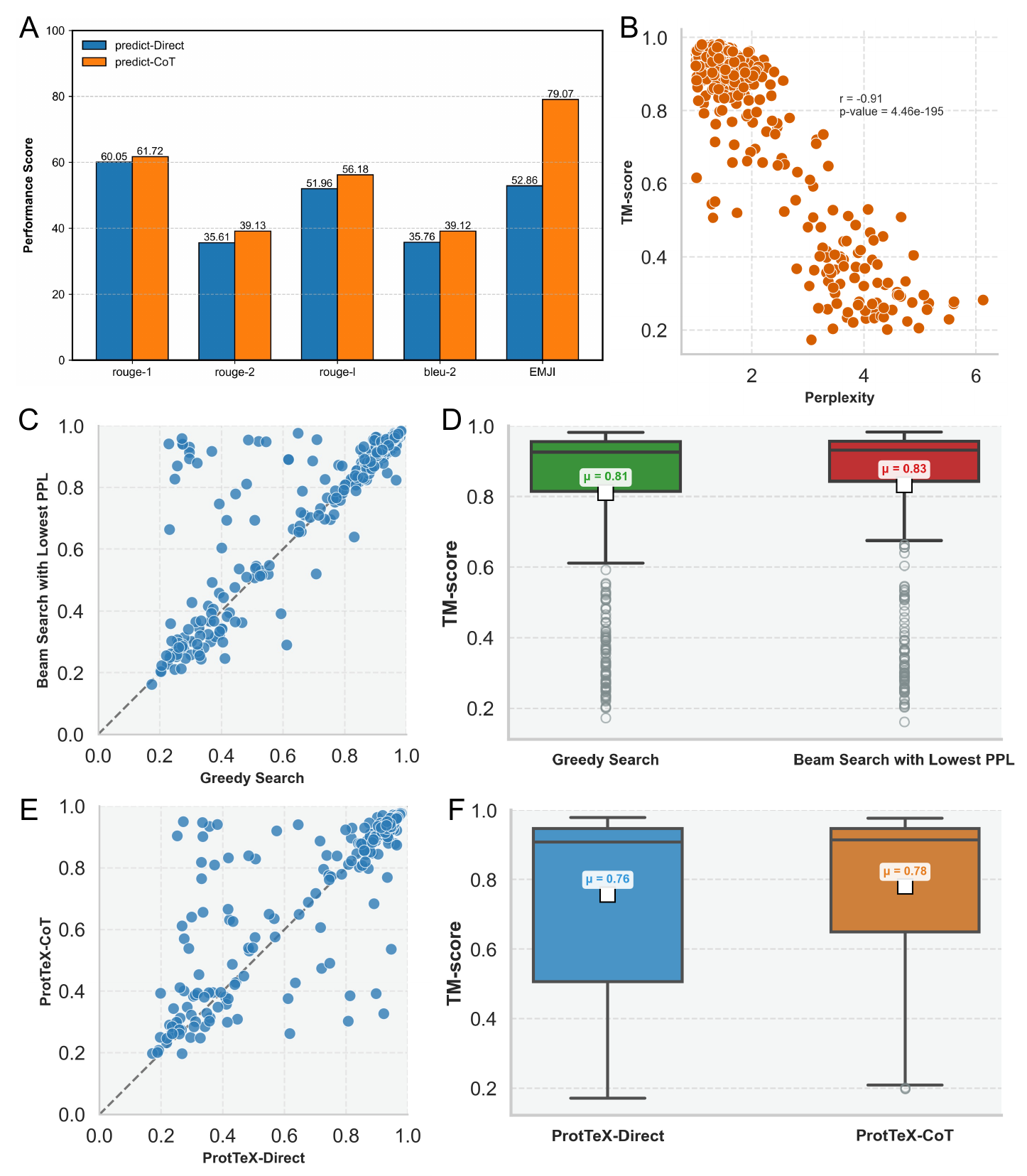} 
  \caption {Multimodal Chain-of-Thought Reasoning performance. (A) Bar plot comparing the performance scores of subcellular location prediction between Direct Prompting and CoT Prompting (n=1978). (B) Scatter plot illustrating the negative correlation between perplexity and TM-score of predicted structures. The Pearson correlation coefficient and corresponding p-value are provided in the legend. (C) \& (D) Comparison of structure prediction performance across Beam Search with Lowest Perplexity and Greedy Search strategies on PSPD test set (n = 500). (E) \& (F) Comparison of structure prediction performance between Direct Prompting and CoT Prompting on PSAD test set (n=500).}
  \label{fig:figure_5}
\end{figure*}

To assess whether incorporating CoT reasoning enhances model performance in functional understanding tasks, we conduct a systematic experiment. We select the subcellular location prediction task in PFUD test set with protein length less than 400 to evaluate our approach. Two distinct prompting strategies were implemented: (1) Direct Prompting: The model directly predicts function from the input protein sequence. (2) CoT Prompting: The model first generates an intermediate reasoning step involving protein structure prediction based on the input sequence, and then utilizes both the original protein sequence and the generated structure to predict function. Results presented in Figure~\ref{fig:figure_5}A demonstrate substantial improvements in model performance when CoT reasoning is employed. The generated outputs exhibit enhanced coherence and improved task alignment, with fewer instances of irrelevant or inconsistent information. Specifically, exact match accuracy improved by 49.6\% compared to the direct prompting approach. These observed improvements suggest that explicit reasoning pathways enable the model to better contextualize multimodal inputs, resulting in a deeper understanding of protein functions. This finding aligns with theoretical CoT frameworks, highlighting the importance of decomposable inference processes in complex prediction tasks. By breaking down the reasoning into intermediate steps, the model can more effectively leverage available information and mitigate errors caused by oversimplified assumptions.

Given the demonstrated effectiveness of Chain-of-Thought (CoT) reasoning in enhancing functional prediction tasks, we further investigate whether an alternative CoT strategy—beginning with sequence analysis followed by structure prediction—can similarly improve performance in structure prediction tasks. Traditional protein structure prediction inherently involves significant computational challenges, typically requiring specialized SE(3)-invariant architectures and resource-intensive, multi-phase training paradigms. In contrast, our model accomplishes this task solely through next-token prediction. Since LLMs are decoder-only generative models, we explored whether the model itself can intrinsically assess the quality of the generated protein structures. To this end, we employed perplexity (PPL)—a widely used metric in natural language processing (NLP)—which measures how effectively a probabilistic model predicts given samples~\cite{bai-etal-2024-longbench}. Empirical validation conducted on 500 randomly selected proteins from the PSPD test set revealed a statistically significant negative correlation between TM-score and output perplexity, as illustrated in Figure~\ref{fig:figure_5}B. This correlation suggests that minimizing perplexity could enhance prediction accuracy. Therefore, we propose a simple sampling strategy termed "Beam Search with Lowest PPL." Detailed descriptions of this method are provided in Section~\ref{sampling}. As demonstrated in Figure~\ref{fig:figure_5}C and Figure~\ref{fig:figure_5}D, our proposed sampling strategy improves structural prediction accuracy, highlighting the effectiveness of sampling-based approaches in enhancing predictive performance. These findings open promising directions for future research aimed at developing more efficient and precise sampling strategies, potentially advancing the application of autoregressive models in multimodal scenarios.

% \begin{figure*}[t]
%  \centering
%   \includegraphics[width=0.6\linewidth]{Figure/negative_corr.png} 
%   \caption {Negative Correlation between PPL and TM-score}
%   \label{fig:PPL}
% \end{figure*}

% \begin{figure*}[t]
%  \centering    
%   \includegraphics[width=0.98\linewidth]{Figure/sample_vs_no.pdf} 
%   \caption {Comparison of Beam Search with Lowest PPL and Greedy Search Sampling on PSPD test set. (A) scatter plot for TM-score in Beam Search with Lowest PPL and Greedy Search. (B) Boxplots for TM-score in Beam Search with Lowest PPL and Greedy Search.}
%   \label{fig:sample_improve}
% \end{figure*}

Next, we explore whether incorporating Chain-of-Thought (CoT) reasoning can further enhance model performance in protein structure prediction tasks. We conducted protein structure CoT generation experiments on a randomly selected subset of 500 proteins from the PSAD test set. The study comparatively analyzed two distinct prompting strategies: (1) Direct Prompting, involving direct structure prediction from input sequences, and (2) CoT Prompting, a multi-stage approach requiring descriptive analysis before structural generation, as illustrated in Figure~\ref{fig:chat}. Our quantitative evaluation, visualized in Figure~\ref{fig:figure_5}E and Figure~\ref{fig:figure_5}F, demonstrates that the CoT prompting paradigm consistently outperforms direct prompting, resulting in improved generation accuracy. This finding underscores the effective cross-modal transferability of reasoning techniques from natural language processing to biological domains. Importantly, the CoT approach not only enhances reasoning accuracy but also reduces the "black-box" nature of the model by introducing transparent reasoning processes. This transparency enables the model to explicitly recognize connections between protein structures and natural language descriptions.

% \begin{figure*}[t]
%   \centering
%   \includegraphics[width=0.98\linewidth]{Figure/COT_vs_Direct.pdf} 
%   \caption {Comparison of CoT Prompting and Direct Prompting on PSAD test set. (A) Scatter plot for TM-score in Direct Prompting and CoT prompting. (B) Boxplots for TM-score in Direct Prompting and CoT Prompting}
%   \label{fig:COT}
% \end{figure*}

Overall, our initial attempt to achieve Chain-of-Thought (CoT) capabilities using only a limited dataset has yielded remarkably promising results. The empirical findings highlight significant potential for practical applications. Our proposed framework addresses the critical challenge of integrating CoT reasoning into biological scientific inference, enabling deeper insights into biological functions and facilitating the rational generation of biologically meaningful molecules.

\subsection{Transaction of Language Decoding Techniques for Protein Structure Sampling}
In LLMs, sampling is a critical step during text generation, with temperature parameters often controlling the balance between accuracy and diversity~\cite{renze-2024-effect}. Similarly, this principle may be beneficial for generating protein structures, where accurately modeling conformational variability is essential. Protein conformational diversity allows proteins to adopt distinct structural states under varying physiological conditions, significantly impacting protein behavior, ligand binding, and allosteric regulation~\cite{henzler-wildmanDynamicPersonalitiesProteins2007}. To explore this concept, we conducted zero-shot experiments to evaluate our model’s capability for generating multiple protein conformations.

We selected a set of nine proteins known to exhibit conformational variability. Specifically, three proteins (KaiB, Mad2, and RfaH) were previously studied by AlphaFold using a multiple sequence alignment (MSA) clustering method~\cite{wayment-steelePredictingMultipleConformations2024c}. The remaining six proteins, MinE, EhCaBP, DDX19, IMPase, Thioesterase, and Capsid Protein, which exhibit co-evolved residue pairs, are classified into Category 1 by W. Schafer et al.~\cite{schaferEvolutionarySelectionProteins2023}. Since our model was not explicitly trained for this task, we defined successful sampling as the ability to sample two distinct conformations, each with a TM-score above 0.7.

\begin{figure*}[t]
  \centering
  \includegraphics[width=0.8\linewidth]{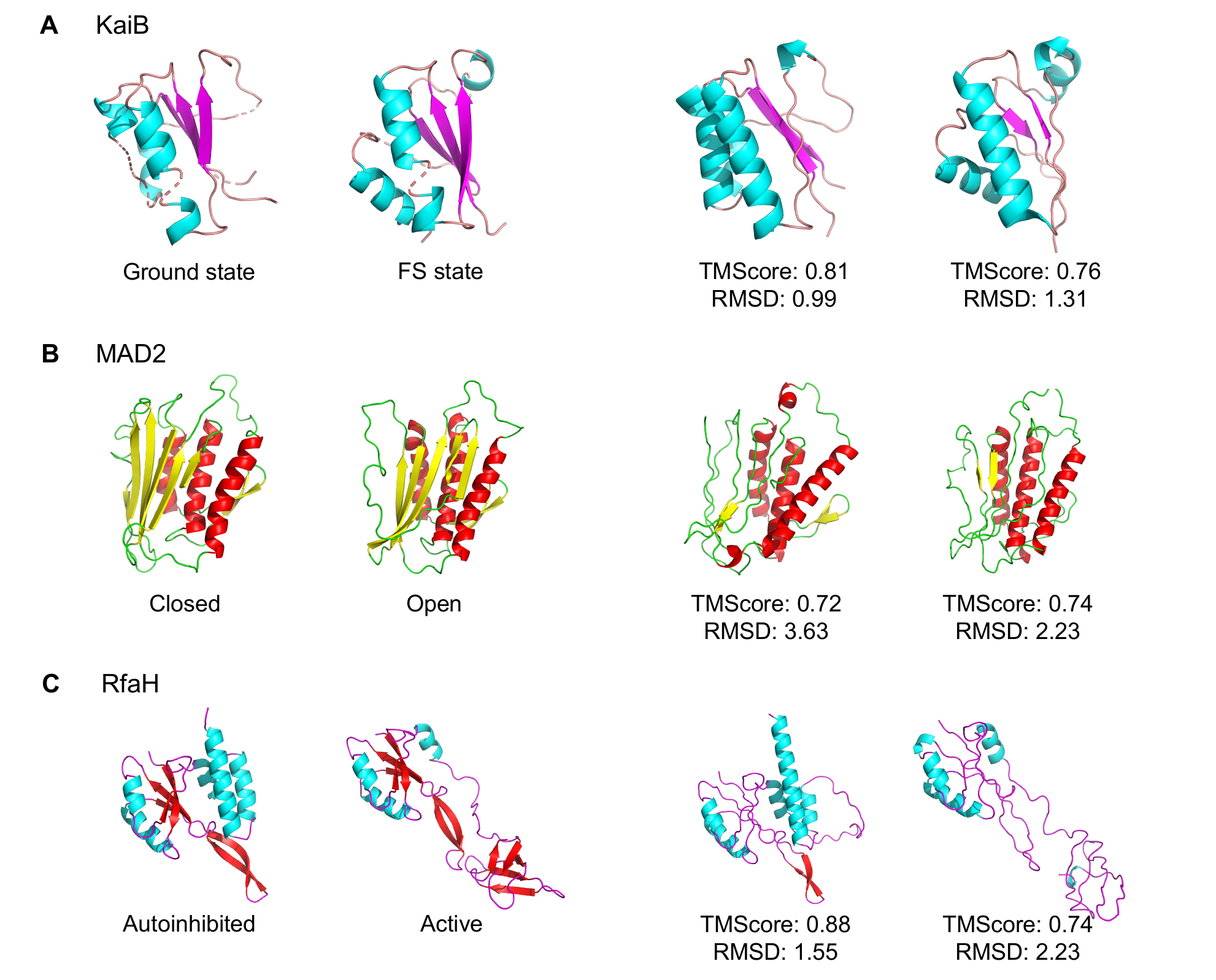} 
  \caption {Multi-conformation sampling for fold-switching proteins (A) KaiB, (B) MAD2 and (C) RfaH}
  \label{fig:multi-main}
\end{figure*}

We utilized the nucleus sampling strategy~\cite{holtzman2020curiouscaseneuraltext}, a widely recognized and effective approach for text generation. Specifically, we set the temperature parameter to 0.7 and the top-p value to 0.4, ensuring controlled generation diversity and quality. For each protein, we generated 100 samples and selected the pairs that exhibitd the highest structural similarity to the two target conformations. Our model successfully sampled 6/9 proteins. As shown in Figure~\ref{fig:multi-main}, all three proteins identified by AFcluster were successfully sampled by our model, although their secondary structures remained suboptimal. The other three successful cases are presented in the Supplementary Figure~\ref{fig:multi-supp}.

During training, the majority of proteins used were relatively stable, which inherently constrained the model’s ability to capture information of multiple conformations. Our preliminary results suggest that evolutionary information embedded within protein sequences can be learned through the mapping between sequence and structural tokens. Furthermore, we have demonstrated that an autoregressive LLM also possesses the potential for conditional protein generation, achieving performance comparable to that of diffusion models. For future work, a systematically curated dataset of multiconformational proteins or an unconditionally generated dataset could be utilized to fine-tune our model. This approach could enhance the model's ability to perform unconditional protein generation or to generate protein structures conditioned on sequence information. Such advancements would contribute to a deeper understanding of protein conformational diversity and its implications in structural biology.

\subsection{Knowledge-guided and natural language-instructed protein design}

\begin{figure*}[t]
  \centering
  \includegraphics[width=0.8\linewidth]{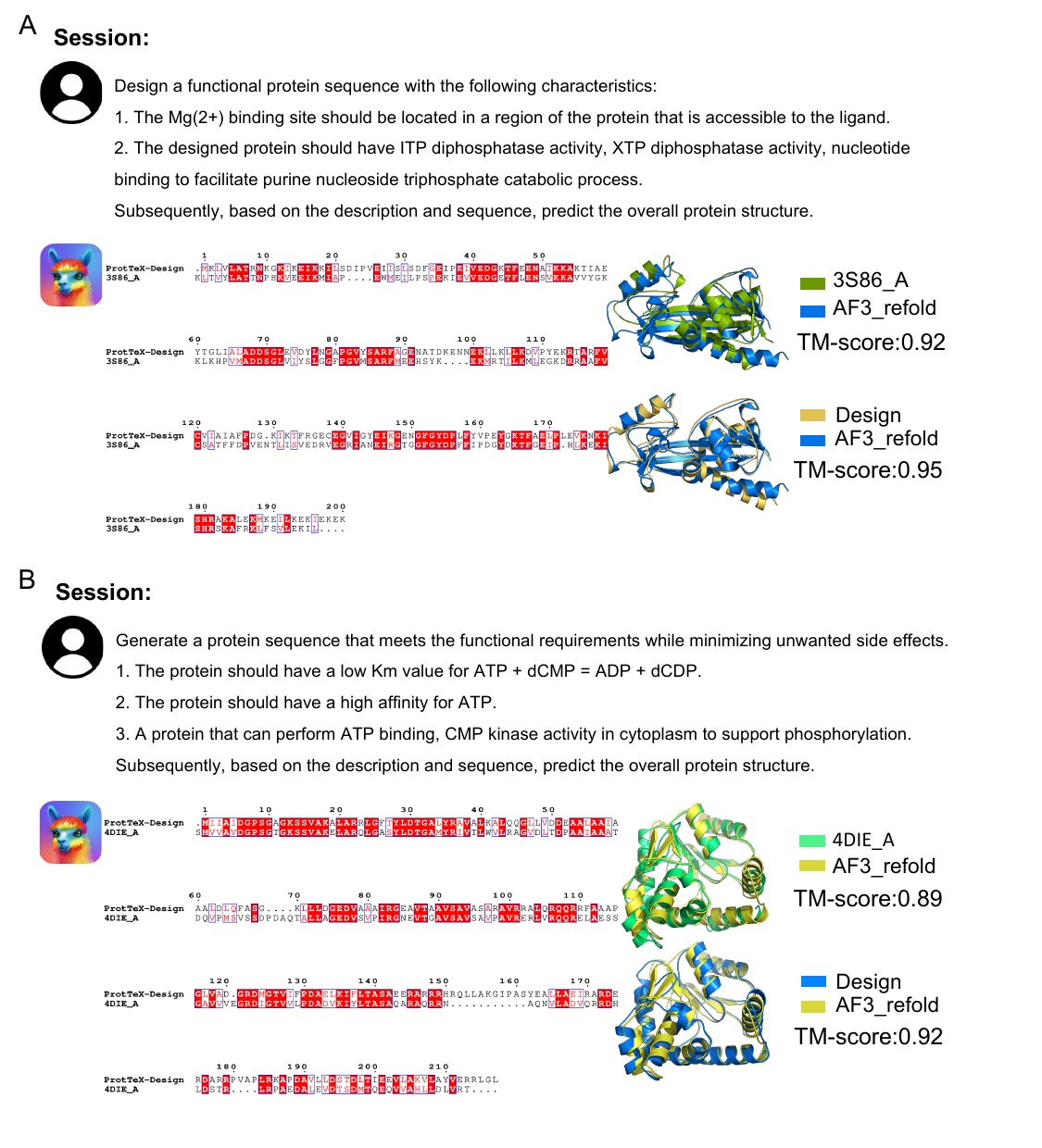} % 调整缩放比例
  \caption{Design chat and results for (A) dITP/XTP pyrophosphatase and (B) cytidylate kinase.}
  \label{fig:design_chat_dXTP}
\end{figure*}

Designing proteins with customizable properties is a long-standing goal in biochemistry. The ability to rapidly and cost-effectively engineer specific, efficient, and tailored proteins holds immense potential for addressing many of the challenges humanity faces today and will encounter in the future. In this study, we investigate the model's capability for controllable protein design, facilitated by the inclusion of a small subset of protein design problems in the Protein Design Dataset (PDD), where we prompt the model to co-generate sequences and structures based on human-designed functional prompts. We perform two case studies in which the model generates protein sequences and structures based on specific functional requirements, as illustrated in Figure~\ref{fig:design_chat_dXTP} 

% \begin{figure*}[t]
%   \centering
%   \includegraphics[width=0.9\linewidth]{Figure/design_cmk_chat.pdf} % 调整缩放比例
%   \caption{Design chat and results for cytidylate kinase}
%   \label{fig:design_chat_cmk}
% \end{figure*}

\begin{figure*}[t]
  \centering
  \includegraphics[width=0.7\linewidth]{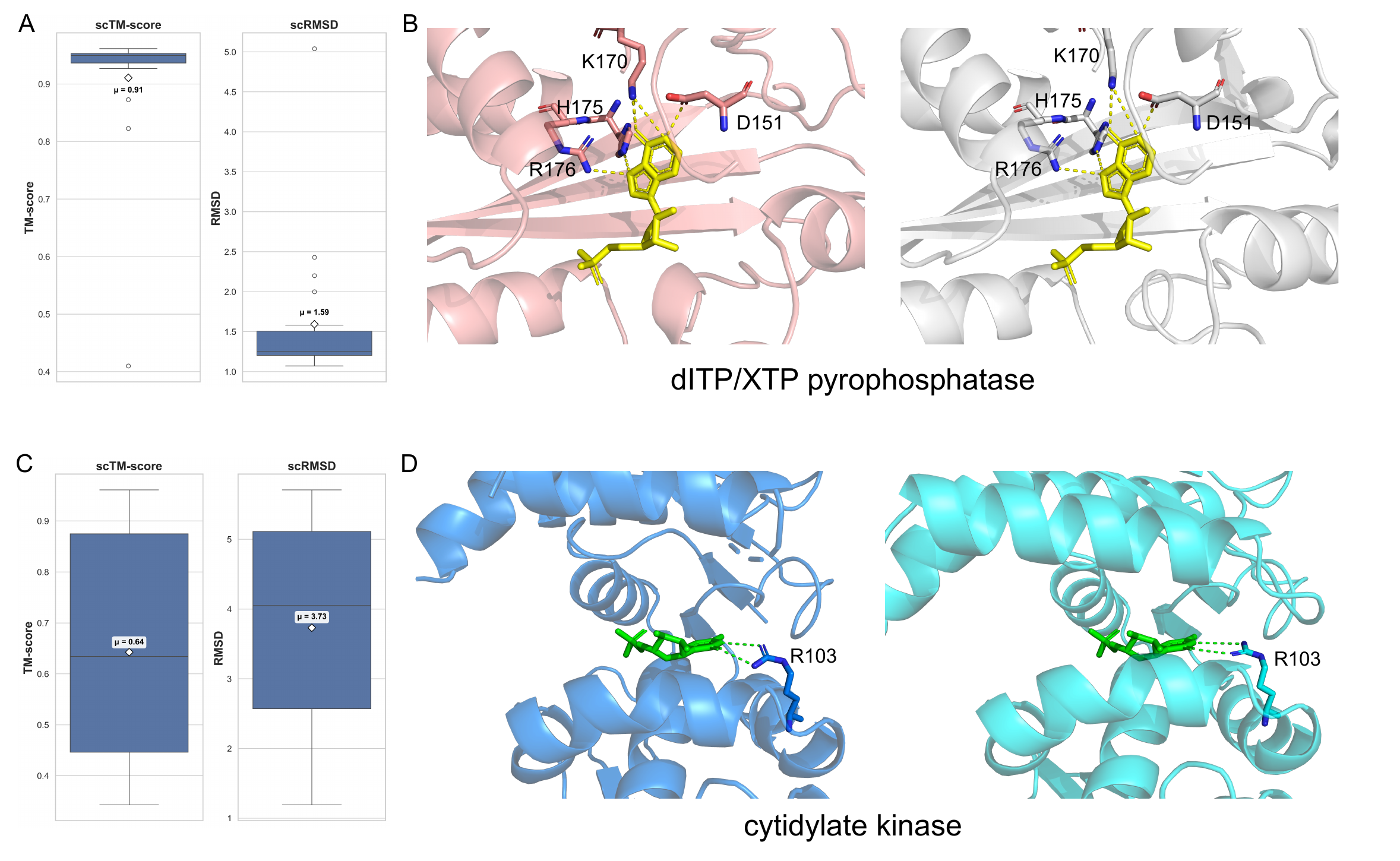} % 调整缩放比例
  \caption{Generation quality for controllable protein design. (A) Self-consistency TM-score and RMSD for dITP/XTP pyrophosphatase. (B) Comparison of the active site between natural (light pink) and designed (gray) dITP/XTP pyrophosphatase, with inosine monophosphate (IMP) highlighted in yellow. (C) Self-consistency TM-score and RMSD for cytidylate kinase. (D) Comparison of the active site between natural (light pink) and designed (gray) cytidylate kinase, with cytidine-5'-monophosphate highlighted in green.}
  \label{fig:generate_quality}
\end{figure*}

The two protein cases we investigated are cytidylate kinase and dITP/XTP pyrophosphatase. Since functional properties such as Mg(2+) binding and ATP binding appear multiple times in our training dataset, the model has effectively learned the structural and functional characteristics that proteins should exhibit. We provided functional prompts to the model and generated 20 protein sequences and structures. We employed the nucleus sampling strategy, setting the temperature to 0.9 and the top-p value to 0.6 to enhance diversity. Our results show that all the generated sequences exhibit sequence similarity scores below 0.8 when compared to the entire training dataset. To further assess self-consistency in folding, we utilized AlphaFold3\cite{abramsonAccurateStructurePrediction2024} to refold the generated sequences. As shown in Figure~\ref{fig:generate_quality}, the designed sequences and structures demonstrate high self-consistency, suggesting their designability. Notably, the generated sequences adopt folding patterns remarkably similar to those of natural enzymes while maintaining low sequence identity. Furthermore, we conducted a comprehensive analysis of the active sites in these structures. Specifically, we obtained the PDB structures of both proteins co-crystallized with small molecules and performed structural alignment with the corresponding structures predicted by AlphaFold3. As illustrated in Figure~\ref{fig:generate_quality}, the side-chain amino acids involved in molecular interactions with specific molecules are well conserved across both designed and natural proteins. This structural conservation strongly suggests that the designed proteins retain their potential catalytic activity, thereby validating the functional integrity of our engineered sequences. Our findings reveal significant potential of our model in the controllable protein design. These results suggest that fine-tuning LLMs could enable them to explicitly comprehend the relationship between human language and biological language. Moving forward, we aim to leverage the power of LLMs to achieve fully controllable protein generation, enabling real-time human-machine interaction and customized protein editing and design, ultimately accelerating the drug development cycle.

\section{Concluding Remarks}
%%期待更多的prompt设计形式以扩充数据, 对蛋白质序列和结构实现定长压缩或许能方便生成.更优的数据配比和分布可能可以让模型效果有更好的提升
In this paper, we present a novel unified framework which achieves dual innovation in protein science. (1) Architectural unification of core challenges through a foundation model with task-agnostic formulation via a single-model, unified-loss paradigm. (2) Pioneering the application of Chain-of-Thought reasoning in the multimodal protein reasoning and generation. ProtTeX enables LLMs to effectively process protein-related tasks through a mixed-modal fusion strategy. By employing in-context supervised fine-tuning, we have successfully integrate domain-specific knowledge into general-purpose LLMs, such as Llama3, equipping them with preliminary capabilities in multimodal protein comprehension and generation. Our model not only demonstrates the ability to tackle a wide range
of protein tasks, including protein understanding, structure generation and protein design, but also introduces multimodal Chain-of-Thought reasoning, enhancing the transparency of the model's deductive processes and ensuring greater interpretability. 

Our current model may exhibit slight performance gaps in certain tasks compared to task-specific or domain-expert models, such as ESMFold\cite{hayesSimulating500Million2025,linEvolutionaryscalePredictionAtomiclevel}. Notably, ESMFold's success is largely attributed to its massive parameter count and extensive training dataset, which surpass even those of Llama3 8B in scale. Considering the well-known scaling laws of LLMs, we can anticipate continuous performance improvements across various tasks with the development of larger LLMs and more biological data. Beyond scaling parameters, well-established paradigms in the LLM domain, such as reinforcement learning-based alignment\cite{Cao_2024} and inference-time self-improvement\cite{dong2024surveyllminferencetimeselfimprovement}, can be systematically applied to enhance ProtTeX’s performance across various protein-related tasks. Exploring these approaches will be a key focus for our future research.


\bibliography{achemso-demo}

\newpage

\appendix

\section{Additional Information}

\subsection{Data and Software Availability}

Our model and dataset will be publicly available on Hugging Face:\url{https://huggingface.co/mzcwd/ProtTeX}

\subsection{Dataset}\label{supp_dataset}

UniProt\cite{10.1093/nar/gkae1010} is a comprehensive and widely used database for protein annotation, enriched with detailed functional information and corresponding protein sequences. Many protein-related question-answering datasets focus on protein functionality. For example, Mol-Instruction\cite{fangMolInstructionsLargeScaleBiomolecular2024a} and ProteinLMBench~\cite{shenFinetuningDatasetBenchmark2024} are both constructed based on the extensive information provided by UniProt.To establish a robust protein structure database, we first construct a dataset based on UniProt. Specifically, we collect the clustered AlphaFold Protein Structure Database (AFDB) v4 dataset\cite{hernandezClusteringPredictedStructures}, which includes 2.27 million single-chain structures predicted before July 25, 2022. Additionally, we extract protein structures from the Swiss-Prot database, filtering and curating the dataset released in May 2022, resulting in 541,327 single-chain structures. Furthermore, we incorporate experimentally determined structures from RCSB PDB\cite{10.1093/nar/28.1.235}, including 551,957 single-chain structures released before October 13, 2021. In total, our dataset comprises 3.36 million protein sequences paired with structural information, forming a comprehensive resource for protein structure analysis and model training.

We then process this dataset using the ProtTeX tokenizer, introduced in Section~\ref{tokenizer}, for structural reconstruction, obtaining a tokenized string for each protein. Proteins with a reconstructed TM-score above 90 are retained. The dataset is subsequently split into training (90\%), validation (5\%), and test (5\%) sets. For all protein-related questions constructed in the following sections, protein accessions are sourced from UniProt. Accordingly, problem-specific datasets are created by matching the corresponding accessions from the training, validation, and test sets. If an accession is not found in our database, the corresponding question is excluded from the dataset.

\textbf{Protein Function Understanding Dataset(PFUD)}. This dataset is derived from Mol-Instruction for benchmarking purposes. Mol-Instruction integrates three instruction systems: molecule-oriented, protein-oriented, and biomolecule-oriented tasks. For our study, we specifically curated the protein-oriented subset. In the original Mol-Instruction dataset, protein-related questions were sourced from both TrEMBL and Swiss-Prot, with TrEMBL accounting for the majority. To ensure the reliability of protein annotations in the training and validation sets, we excluded TrEMBL entries (which contain unverified annotations) and augmented the Swiss-Prot subset using similar question templates. The final dataset comprises 429,201 proteins from Swiss-Prot. Using Swiss-Prot accession numbers as anchors, we construct both single-turn and multi-turn dialogue datasets. These datasets cover a diverse range of questions spanning multiple domains, including protein feature overviews, recognition of protein domains or motifs, biological processes, molecular functions, subcellular localization, and multi-attribute queries. The "feature overview" category requires the model to generate a concise yet comprehensive summary of a protein’s primary functional characteristics, often integrating information from both molecular function and biological processes. The "multi-attribute" category presents prompts that combine two or three attributes from the aforementioned categories, enabling a holistic analysis of the protein’s properties. For benchmarking protein function understanding, we utilize 5,836 single-turn chat items from the test set. The final dataset comprises 404,640 samples for training, 16,859 samples for validation, and 7,702 samples for testing.

\textbf{Protein Structure Analysis Dataset(PSAD)}. We curate the dataset from proteinLMBench~\cite{shenFinetuningDatasetBenchmark2024}. The original questions required the model to generate descriptions of protein subunit composition, including the organization and interactions between subunits based on the given protein sequence. To enhance the model’s capabilities, we redesigned the prompt to enable protein structure generation following the completion of structural analysis, as illustrated in Figure~\ref{fig:model}C. This modification allows the model to achieve multimodal Chain-of-Thought (CoT) functionality.

\textbf{Protein Design dataset(PDD)}. This dataset is curated from the original Mol-Instruction dataset for protein design. While the original prompts focused on generating protein sequences based on functional descriptions, we have extended the prompt design to predict protein structures after sequence generation, thereby enabling multimodal CoT functionality. Due to the limited availability of data in this subset, we retained sequences whose accession numbers do not appear in our total dataset, restricting them to sequence generation tasks only, without requiring structure prediction. Currently, the dataset size remains relatively small. However, our model architecture is designed to support fine-tuning in this domain, allowing for enhanced performance as additional data becomes available in future work.

%%% protein dit: 只考虑afdb和pdb， 用plddt + reconstruction > 0.95 过滤做了ablation数据集产生760000的数据集，小型数据集做ablation  structure predict-full structure-pred small
\textbf{Protein Structure Prediction Dataset(PSPD)}. A substantial volume of data mapping protein sequences to structures is incorporated, enabling the model to effectively learn the relationship between sequences and their corresponding structures. To ensure the integrity of the training data, we exclude protein accessions that appeared in the training process of the three previously mentioned datasets, utilizing only the remaining data for this training subset.

The total dataset for Proteleon training comprises four key components: PFUD, PSAD, PDD, and PSPD. To ensure robust training and mitigate potential biases introduced by data order, the dataset is randomly shuffled at the beginning of each epoch. The token counts for each subset are presented in Table~\ref{tab:dataset}.

\subsection{Training Details}\label{training_details}
In our experiments, we employ both full-parameter fine-tuning and Low-Rank Adaptation (LoRA)~\cite{hu2021loralowrankadaptationlarge} to optimize the performance of LLMs. To determine the most effective approach, we conduct ablation studies, which reveal that LoRA significantly underperforms compared to full-parameter fine-tuning in terms of task-specific metrics. Consequently, we adopt full-parameter fine-tuning as our primary training strategy. For optimization, we use the AdamW optimizer with a weight decay of 0.1. During training, we apply a cosine annealing learning rate scheduler, gradually decreasing the learning rate from 5e-6 to 1e-7, complemented by a warm-up phase with a ratio of 0.01. Training is conducted on the entire dataset with a batch size of 3 per device over 4 epochs, requiring approximately 5 days to complete on a cluster of 16 NVIDIA A100 GPUs.

\subsection{Ablation details}\label{supp_ablation}

As shown in Table~\ref{tab:result_ablation}, we conduct a systematic ablation study on protein understanding tasks, examining the effects of different training datasets, training strategies, and model scales.

\textbf{Ablation on different training data}. To evaluate the impact of different training strategies, we conduct our ablation study on the PFUD dataset. We only fine-tune the base model using PFUD dataset. Furthermore, we investigate whether sequence-structure pretraining enhances functional understanding by conducting ablation experiments on continued pretraining strategies. As detailed in Table~\ref{tab:result_ablation} and Supplementary Figure~\ref{fig:supp_loss}, our analysis reveals that even a single epoch of pretraining significantly improves model performance on functional comprehension tasks, as shown in Table~\ref{tab:result_ablation}. This finding suggests that enhanced token representations provide critical inductive biases for biological function prediction. Moreover, it further demonstrates that different tasks can mutually reinforce each other, even in the absence of explicit correlations between them. Our final model is unified supervised fine-tunned across multiple downstream tasks, including PSPD. We do not perform additional pretraining on PSPD, as pretraining and fine-tuning would introduce redundancy.

\textbf{Ablation on training strategy}. We compare the LoRA (Low-Rank Adaptation)~\cite{hu2021loralowrankadaptationlarge} training method with full-parameter fine-tuning. For full-parameter fine-tuning, we adopt the configuration previously described in Section~\ref{training_details}. For LoRA implementation, we configure the following parameters: LoRA target set to `all', LoRA rank = 8, and alpha = 16. The learning rate is set higher than that of full-parameter fine-tuning, initialized at 1e-4 and gradually decayed to 1e-6, while all other hyperparameters remain consistent with the full-parameter fine-tuning setup. Our experimental results indicate that LoRA training exhibited significantly slower convergence in validation loss compared to full-parameter fine-tuning, as illustrated in the supplementary figures~\ref{fig:supp_loss}. Moreover, with the application of DeepSpeed Zero Stage 3 technology, LoRA does not demonstrate significant advantages in terms of memory usage or training speed across multiple GPUs compared to full-parameter fine-tuning. Consequently, we adopt full-parameter fine-tuning as our standard training paradigm in subsequent experiments. 

\textbf{Ablation on model scaling}. We conduct ablation experiments to evaluate the impact of different model sizes. The results indicate that the 8B-parameter model slightly outperforms the 1B-parameter model on the given task. However, the performance gap remains relatively small, which is not unexpected given the limited dataset size. Considering the expanded training dataset and the computational scalability requirements for downstream applications, we ultimately select the 8B-parameter model as our foundational architecture.

\setcounter{figure}{0}
\renewcommand{\thefigure}{S\arabic{figure}}

\begin{figure*}[ht]
  \centering
  \includegraphics[width=0.8\linewidth]{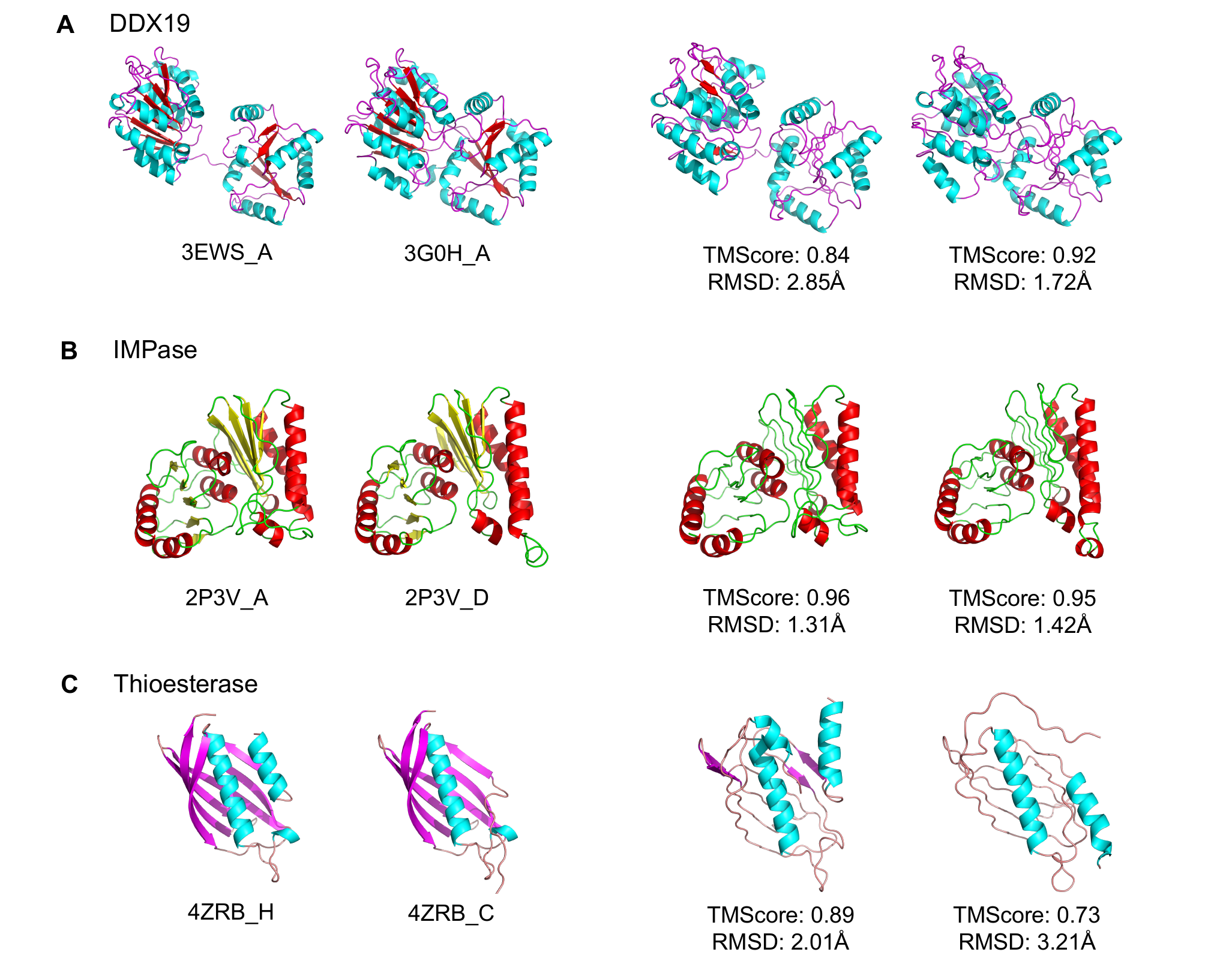} 
  \caption {Multi-conformation sampling for fold-switching proteins in W. Schafer et al.~\cite{schaferEvolutionarySelectionProteins2023}.}
  \label{fig:multi-supp}
\end{figure*}

\begin{figure*}[ht]
  \centering
  \includegraphics[width=0.7\linewidth]{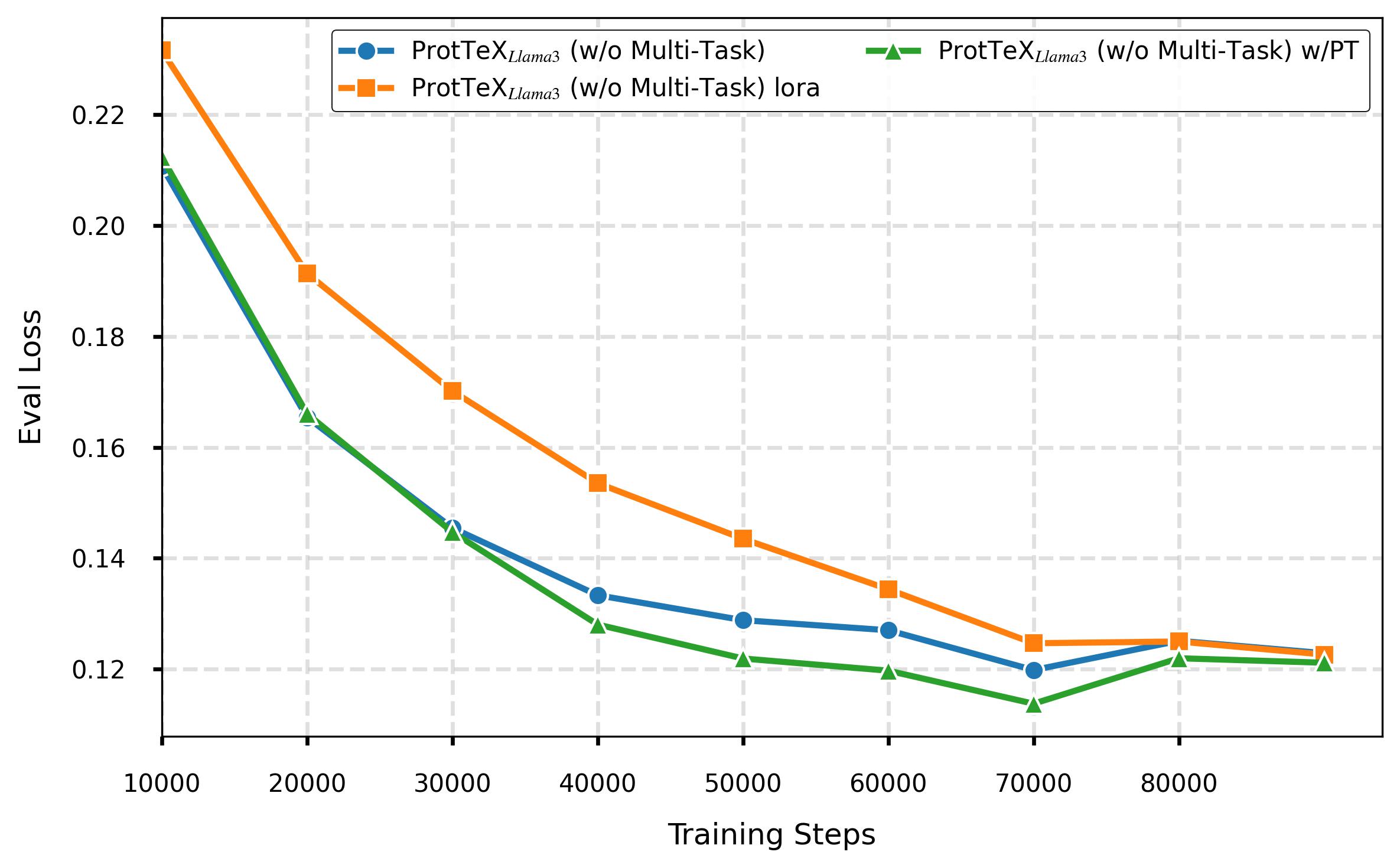} % 调整缩放比例
  \caption{Evaluation loss of different training strategy}
  \label{fig:supp_loss}
\end{figure*}

\end{document}